# Relativistic integro-differential form of the Lorentz-Dirac equation in 3D without runaways

*Michael Ibison, Harold E. Puthoff*

Institute for Advanced Studies at Austin

4030 Braker Lane West, Suite 300 Austin, TX 78759, USA

ibison@earthtech.org, puthoff@aol.com

**Abstract**

It is well known that the third-order Lorentz-Dirac equation admits 'runaway' solutions wherein the energy of the particle grows without limit, even when there is no external force. These solutions can be denied simply on physical grounds, and on the basis of careful analysis of the correspondence between classical and quantum theory. Nonetheless, one would prefer an equation that did not admit unphysical behavior at the outset. Such an equation - an integro-differential version of the Lorentz-Dirac equation – is currently available either in 1 dimension only, or in 3 dimensions only in the non-relativistic limit.

It is shown herein how the Lorentz-Dirac equation may be integrated without approximation, and is thereby converted to a second-order integro-differential equation in 3D satisfying the above requirement. I.E., as a result, no additional constraints on the solutions are required because runaway solutions are intrinsically absent. The derivation is placed within the historical context established by standard works on classical electrodynamics by Rohrlich, and by Jackson.

**Introduction**

The Lorentz-Dirac equation (LDE) describes the motion of a classical charged particle subject to both an external force and self-interaction due to radiation. An undesirable characteristic is the prediction of an exponential (runaway) acceleration in the absence of an applied force. The source of the trouble may be traced to the third order derivative with respect to time. Since one would prefer a second order equation anyhow, a natural approach is to convert the original LDE into a second order equation by integrating over time. At the same time, one might take the opportunity to eliminate the runaway solution by a suitable choice for the constant of integration.

This is the method cited by Jackson [1], as it applies to a non-relativistic (and thereby linearized) version of the LDE. It is successful in that runaway solutions are absent. The same approach was employed by Rohrlich [2] to the relativistic LDE, but without success; his resulting equation still permits runaway solutions. The attempt failed because he was unable to combine the acceleration and radiation parts (times an integrating factor) as a total differential in proper time.

Jackson and Rohrlich are referred to herein because they are standard texts on classical theory. However, for an earlier review of the subject that is both lucid and thorough, the reader is referred to Erber [3]. The first appearance of the non-relativistic integro-differential form of the LDE is due to Haag [4], (and subsequently - in English - Plass [5]).

It has been shown that the non-relativistic integro-differential form of the LDE is the finite-point limit of a finite-size (non-relativistic) model of the electron [6, 7, 8]. Since the latter is free of runaway solutions, this may be regarded as evidence in favor of the validity of the integro-differential form, over the original LDE. Also, (very importantly), Sharp [9] has shown that the non-relativistic integro-differential LDE corresponds to the quantum theory of a non-relativistic charge coupled to the quantized electromagnetic field (neither of which, therefore, display runaway solutions). Both these results point to the need for a relativistic generalization of the existing non-relativistic integro-differential version of the LDE.

Barut [10] has proposed a method to eliminate both the runaway and pre-acceleration behavior of the LDE by making the Abraham vector disappear when the external field disappears. However, as pointed out by Blanco [11], such an equation is essentially quite different from the original LDE. Jimenez and Hirsch [12] suggest that the non-relativistic LDE be supplemented by an external, stochastic, electromagnetic field, in the spirit of Stochastic Electrodynamics (see for instance [13])). This, they argue, has the effect of eliminating the undesirable runaway behavior without modification of the LDE (to an integro-differential form). Their program, though promising, potentially suffers from an externally-induced runaway problem unless the stochastic field is band-limited (which would be similar to supposing a finite-sized charge).

**Runaway solutions of the Lorentz-Dirac equation**

The Lorentz-Dirac equation in proper time is [1,2]

$$m_0 a - m_0 \tau_0 \left( \frac{da}{d\tau} + a^2 u \right) = f(\tau), \qquad (1)$$

where the force $f$ can depend on $\tau$ explicitly, and implicitly via the position and its derivatives. $a \equiv \{a^\mu\} \equiv \{a_0, \mathbf{a}\}$ is the proper acceleration, $u \equiv \{u^\mu\} \equiv \{u_0, \mathbf{u}\} = \{\gamma, \mathbf{u}\}$ is the proper velocity, and $a^2 = (a_0)^2 - \mathbf{a}.\mathbf{a}$, $c = 1$, and $\tau_0 = e^2/6\pi\varepsilon_0 m_0$ is (2/3) the time it takes for light to travel across the classical electron radius. The notorious runaway solution is most easily demonstrated in one dimension, wherein the LDE is easily linearized [2]. With the substitution $dx/d\tau = \sinh(w(\tau))$, one obtains from Eq. (1)

$$\dot{w} - \tau_0 \ddot{w} = f/m_0 , \qquad (2)$$

where $f$ is the ordinary Newton force in the $x$ direction. It is clear that even when there is no external force, ($f = 0$), $w$ may increase without limit, since $w \sim \exp(\tau/\tau_0)$ is a solution. This causes $dx/d\tau$ and $\gamma$ to increase without limit, giving rise to the interpretation that the particle has accelerated to the speed of light and has acquired an infinite kinetic energy. The reason for the presence of such solutions may be traced to the intrinsically non-conservative nature of the equation of motion. It was conceived to account for losses due to radiation, but turns out to admit gains, presumably by the same mechanism.

**The non-relativistic integro-differential equation**

When the velocities are small compared to $c$, $\gamma \approx 1$, $d\tau \approx dt$, and Eq. (1) becomes

$$m_0 \mathbf{a} - m_0 \tau_0 \frac{d\mathbf{a}}{dt} = \mathbf{f} . \qquad (3)$$

(This non-relativistic form of the LDE is also called the Abraham-Lorentz equation.) It suffers from the same runaway solution as Eq. (2) - the relativistic one-dimensional result written in hyperbolic co-ordinates. The traditional remedy [1] is to replace Eq. (3) with the integro-differential equation

$$m_0 \mathbf{a} = \int_0^\infty ds\, e^{-s} \mathbf{f}(t + s\tau_0) . \qquad (4)$$

It is readily verified upon substitution that the **x** that solve this equation are a subset of those that solve Eq. (3). It is also clear that, provided **f** vanishes in the remote future, the acceleration also vanishes in the remote future. Not only does this prescription eliminate the runaway solution, but it also restores the boundary condition requirements to those of a second order differential equation, e.g.: the position and velocity are given at some time. This time need not be when the force is zero (i.e. the remote past or the remote future); it may be any time. Though the runaway behavior is tamed, it is at the expense of an acausal connection between the applied force and the resulting acceleration. Specifically, it is seen from Eq. (4) that the acceleration depends on *future* forces (exhibits pre-acceleration). However, the temporal range, $\tau_0$, of that dependency, is such that pre-acceleration is too small to be observed on classical time scales.

**Rohrlich's relativistic integro-differential equation**

It is carefully argued by Rohrlich [2] that runaway solutions must be denied by imposing a suitable constraint, i.e., a boundary condition on the acceleration. In this paper, we will be content with the condition

$$\lim_{\tau \to +\infty} \frac{d^2 x^\mu}{d\tau^2} = 0 \Leftrightarrow \lim_{t \to +\infty} \frac{d^2 x^\mu}{dt^2} = 0 \Leftrightarrow \lim_{t \to +\infty} \frac{d^2 \mathbf{x}}{dt^2} = 0, \tag{5}$$

since we require an acceptable prediction of future behavior based on some 'initial' condition, given at some nominal but finite time. With the aim of integrating the constraint into the equation of motion, Rohrlich investigates a formal integration of Eq. (1),

$$a_\mu = A_\mu e^{\tau/\tau_0} + \int_\tau^\infty d\tau' e^{(\tau-\tau')/\tau_0} \left( \frac{1}{m_0 \tau_0} f_\mu(\tau') + a^2(\tau') u_\mu(\tau') \right), \tag{6}$$

where $A_\mu$ is a 4-vector constant of integration. He sets $A_\mu = 0$, and considers the new equation as a possible replacement for Eq. (1). However, as he points out, setting $A_\mu$ to zero guarantees only that $\lim_{\tau \to \infty} e^{-\tau/\tau_0} a_\mu = 0$ which, clearly, is weaker than the requirement that the acceleration vanish, Eq. (5). Therefore we conclude that Eq. (6) with $A_\mu = 0$ is unsatisfactory, since a supplemental constraint must still be imposed to filter out the unphysical behavior.

**An integrating factor for the Lorentz-Dirac equation**

A fully relativistic integro-differential form of the Lorentz-Dirac equation that does not admit runaway solutions (and therefore does not require supplemental constraints) is possible if a suitable integrating factor for the original LDE can be found. If it exists, an integrating factor $S \equiv \{S_\mu{}^\nu(\tau)\}$ satisfying

$$\frac{d}{d\tau}(Sa) = -\frac{1}{\tau_0 m_0} Sf, \tag{7}$$

will permit - via the integration of Eq. (7) – the imposition of boundary conditions Eq. (5) on the acceleration. For this integrating factor to exist, by carrying out the differentiation in Eq. (7) and comparing with Eq. (1) left multiplied by $S$, it must be true that

$$\left(\frac{dS}{d\tau} + \frac{S}{\tau_0}\right)a = a^2 Su,\qquad(8)$$

where none of the elements of $S$ can depend on the acceleration $a$. A substitution into Eq. (8) of

$$S = Re^{-\tau/\tau_0}\qquad(9)$$

where $R \equiv \{R_\mu{}^\nu(\tau)\}$, removes the exponential decay factor to give the requirement that $R$ satisfy

$$\frac{dR}{d\tau}a = a^2 Ru\,.\qquad(10)$$

There are only three independent equations in Eq. (1) because the product of both sides with the four-velocity is identically zero. As a consequence, for any $b_\mu(\tau)$, $R_\mu{}^\nu = b_\mu(\tau)u^\nu$ sets each side of Eq. (7) to zero, and so cannot be a candidate for the integrating factor. It follows that $R$ cannot have a unique solution, since any candidate solution $R_\mu{}^\nu = C_\mu{}^\nu$ say will generate a family of solutions just by addition of this 'null' solution: $R_\mu{}^\nu = C_\mu{}^\nu + b_\mu(\tau)u^\nu$. Of course, whatever form is chosen, that choice cannot impact the equation of motion for each component of $x_\mu$.

With the sign convention $\{q_\mu\} \equiv \{q_0, -\mathbf{q}\}$, a particularly simple solution of Eq. (1) for the integrating factor is

$$R = \{R_\mu{}^\nu\} = \begin{pmatrix} u_0 & u_1 & u_2 & u_3 \\ u_1 & u_0 & 0 & 0 \\ u_2 & 0 & u_0 & 0 \\ u_3 & 0 & 0 & u_0 \end{pmatrix} = \{c_\mu u^\nu - u_\mu c^\nu + u_0 \delta_\mu{}^\nu\}\qquad(11)$$

where $\{c_\mu\} \equiv \{1,0,0,0\}$ is a unit time-like vector. With this definition, one easily sees that Eq. (10) is satisfied, and in particular that the two terms are

$$\frac{dR_\mu{}^\nu}{d\tau}a_\nu = a^2 R_\mu{}^\nu u_\nu = a^2 c_\mu\,.\qquad(12)$$

Recalling Eq.(9), it follows that the Lorentz-Dirac equation, Eq. (1), may be written

$$\frac{d}{d\tau}\left(e^{-\tau/\tau_0}Ra\right) = -\frac{e^{-\tau/\tau_0}}{\tau_0 m_0}Rf\,,\qquad(13)$$

where $R$ is given by Eq. (11), and the inverse of $R$, denoted here by $\hat{R}$, is

$$R^{-1} \equiv \{\hat{R}_\mu{}^\nu\} = \frac{1}{u_0}\begin{pmatrix} u_0^2 & -u_0u_1 & -u_0u_2 & -u_0u_3 \\ -u_0u_1 & u_1^2+1 & u_1u_2 & u_1u_3 \\ -u_0u_2 & u_1u_2 & u_2^2+1 & u_2u_3 \\ -u_0u_3 & u_1u_3 & u_2u_3 & u_3^2+1 \end{pmatrix} = \left\{\frac{1}{u_0}\left(\delta_\mu{}^\nu - u_\mu u^\nu - c_\mu c^\nu\right) + 2u_\mu c^\nu\right\}. \tag{14}$$

$R$ does not behave like a tensor under boosts, and is therefore not a Lorentz tensor. However, it does behave like a tensor under spatial rotations and space and time translations, and is therefore a Euclidean tensor. Nonetheless, the Lorentz invariance of the Lorentz-Dirac equation is preserved. This can be seen more readily if Eq. (13) is written as

$$m_0 a - m_0 \tau_0 \left(\frac{da}{d\tau} + \hat{R}\frac{dR}{d\tau}a\right) = f, \tag{15}$$

whereupon it apparent that the requirement is not that $R$ be a Lorentz tensor, but that $\hat{R}\dfrac{dR}{d\tau}a$ be a true 4-vector. The latter is guaranteed by design. Specifically it is equal to $a^2 u$, in conformity with Eq. (1), as may be confirmed using Eqs. (11) and (14).

**Integration and imposition of the boundary condition**

Formally, the first integral of Eq. (13) is

$$e^{-\tau/\tau_0}R(\tau)a(\tau) - e^{-\tau_c/\tau_0}R(\tau_c)a(\tau_c) = -\frac{1}{\tau_0 m_0}\int_{\tau_c}^{\tau}d\tau' e^{-\tau'/\tau_0}R(\tau')f(\tau')$$
$$\Rightarrow a(\tau) = e^{\frac{\tau-\tau_c}{\tau_0}}R^{-1}(\tau)R(\tau_c)a(\tau_c) - \frac{1}{\tau_0 m_0}\int_{\tau_c}^{\tau}d\tau' e^{\frac{\tau-\tau'}{\tau_0}}R^{-1}(\tau)R(\tau')f(\tau') \tag{16}$$

where $\tau_c$ is the time at which the proper acceleration is presumed known. We are now in a position to impose the requirement that the acceleration in the remote future, $\tau_c = +\infty$ - when the force has long since vanished - is zero. With $a(\tau_c) = 0$, Eq. (16) becomes

$$a(\tau) = \frac{1}{\tau_0 m_0}\int_{\tau}^{\infty}d\tau' e^{\frac{\tau-\tau'}{\tau_0}}R^{-1}(\tau)R(\tau')f(\tau'). \tag{17}$$

Upon the change of variable $s = (\tau' - \tau)/\tau_0$, this is

$$m_0 a(\tau) = \int_0^\infty ds\, e^{-s}R^{-1}(\tau)R(\tau+s\tau_0)f(\tau+s\tau_0) \tag{18}$$

which may be recognized as a relativistic version of the non-relativistic form, Eq. (4). It is easily seen that, having isolated the second derivative on the left hand side, the acceleration is guaranteed to vanish in the remote future if the force also vanishes then. Therefore, the solution is evidently free of runaways. Further, it is evident that solutions of this equation are a subset of the solutions of the original Lorentz-Dirac equation, Eq. (1). Therefore, it can be concluded that the integro-differential equation Eq. (18) is the physically correct equation of motion for a classical charged particle; it retains the properties of the original Lorentz-Dirac equation without the unphysical behavior.

Since it is not immediately evident from Eq. (18), we here confirm that, as required, the acceleration is orthogonal to the velocity. Taking the 4-vector product of Eq. (18) with the velocity gives

$$u^{\mu}(\tau)a_{\mu}(\tau) = \int_{0}^{\infty} ds\, e^{-s} u^{\mu}(\tau) \hat{R}_{\mu}^{\ \nu}(\tau) R_{\nu}^{\ \lambda}(\tau+s\tau_0) f_{\lambda}(\tau+s\tau_0). \tag{19}$$

Using Eq. (14) one finds that

$$u^{\mu} \hat{R}_{\mu}^{\ \nu} = u^{\mu} \left( \frac{1}{u_0}\left(\delta_{\mu}^{\ \nu} - u_{\mu}u^{\nu} - c_{\mu}c^{\nu}\right) + 2u_{\mu}c^{\nu} \right) = c^{\nu}. \tag{20}$$

Inserting this into Eq. (19) and then using Eq. (11) gives

$$u^{\mu}(\tau)a_{\mu}(\tau) = \int_{0}^{\infty} ds\, e^{-s} R_0^{\ \lambda}(\tau+s\tau_0) f_{\lambda}(\tau+s\tau_0) = \int_{0}^{\infty} ds\, e^{-s} u^{\lambda}(\tau+s\tau_0) f_{\lambda}(\tau+s\tau_0) = 0, \tag{21}$$

where the last step follows because the 4-force is required to be orthogonal to the velocity.

**Proper-time vector form**

The 3-vector form of Eq. (18) is obtained as follows. Given $f_{\lambda} = \{\mathbf{u}.\mathbf{f}, -u_0 \mathbf{f}\}$, where $\mathbf{f}$ is the ordinary Newton force vector (i.e., borrowed from $d\mathbf{p}/dt = \mathbf{f}$), and $u_0 = \gamma = \sqrt{1+\mathbf{u}.\mathbf{u}}$, then, using Eq. (11), one obtains

$$R_{\nu}^{\ \lambda} f_{\lambda} = \left(c_{\nu} u^{\lambda} - u_{\nu} c^{\lambda} + u_0 \delta_{\nu}^{\ \lambda}\right) f_{\lambda} = -u_{\nu} f_0 + u_0 f_{\nu} = \{0, (\mathbf{u}\mathbf{u}^T - u_0^2)\mathbf{f}\} = \{0, \mathbf{u}\times(\mathbf{u}\times\mathbf{f}) - \mathbf{f}\}. \tag{22}$$

Denoting the three-space part by $\mathbf{w} \equiv \mathbf{u}\times(\mathbf{u}\times\mathbf{f}) - \mathbf{f}$, Eq. (18) can be written

$$m_0 \mathbf{a}(\tau) = -\sup_{3\times 3}\{R^{-1}(\tau)\} \int_{0}^{\infty} ds\, e^{-s} \mathbf{w}(\tau+s\tau_0) \tag{23}$$

where $\boldsymbol{\alpha}$ is the proper acceleration, and where the sub operation extracts the 3x3 (spatial) sub-matrix. Using Eq. (14) the latter is easily seen to be

$$\mathop{\text{sub}}_{3\text{x}3}\{R^{-1}\} = \frac{1}{u_0}\left(1+\mathbf{u}\mathbf{u}^T\right) \tag{24}$$

whereupon Eq. (23) gives the integro-differential version of the LDE in proper-time vector form:

$$m_0\boldsymbol{\alpha} = \gamma^{-1}\left(1+\mathbf{u}\mathbf{u}^T\right)\int_0^\infty ds\, e^{-s}\left(\gamma^2 - \mathbf{u}\mathbf{u}^T\right)\mathbf{f} = \gamma^{-1}\left(1+\mathbf{u}\mathbf{u}^T\right)\int_0^\infty ds\, e^{-s}\left(\mathbf{f}-\mathbf{u}\times(\mathbf{u}\times\mathbf{f})\right), \tag{25}$$

where the functions in the integrand are to be evaluated at $\tau + s\tau_0$. In particular, if $\mathbf{f}$ is the Lorentz force, $\mathbf{f} = e(\mathbf{E}+\mathbf{u}\times\mathbf{B}/\gamma)$, then the proper acceleration is

$$m_0\boldsymbol{\alpha} = e\gamma^{-1}\left(1+\mathbf{u}\mathbf{u}^T\right)\int_0^\infty ds\, e^{-s}\left(\mathbf{E}-\mathbf{u}\times(\mathbf{u}\times\mathbf{E})+\gamma\mathbf{u}\times\mathbf{B}\right). \tag{26}$$

To write the proper acceleration in terms of vector cross-products, it is useful to define an intermediate quantity

$$\overline{\mathbf{f}} \equiv \int_0^\infty ds\, e^{-s}\left(\mathbf{f}-\mathbf{u}\times(\mathbf{u}\times\mathbf{f})\right), \tag{27}$$

where once again the functions in the integrand are to be evaluated at $\tau + s\tau_0$. With this substitution, an alternative form for Eq. (26) is therefore

$$m_0\boldsymbol{\alpha} = \gamma\overline{\mathbf{f}} + \mathbf{u}\times\left(\mathbf{u}\times\overline{\mathbf{f}}\right)/\gamma. \tag{28}$$

**Proper-time series expansion in $\tau_0$**

A series expansion of the integrand in ascending powers of $\tau_0$ can be expected to converge rapidly if the projection of the force - $\left(\gamma^2 - \mathbf{u}\mathbf{u}^T\right)\mathbf{f}$ - is slowly varying on the time scale of the classical time $\tau_0$. From Eq. (18), one has

$$m_0 a = R^{-1}\sum_{n=0}^\infty \left(\tau_0\frac{d}{d\tau}\right)^n (Rf) \tag{29}$$

where all functions are now evaluated at time $\tau$. In vector form this is

$$m_0\boldsymbol{\alpha} = \gamma^{-1}\left(1+\mathbf{u}\mathbf{u}^T\right)\sum_{n=0}^\infty \left(\tau_0\frac{d}{d\tau}\right)^n \left(\gamma^2-\mathbf{u}\mathbf{u}^T\right)\mathbf{f}. \tag{30}$$

**Ordinary-time vector form**

The integro-differential form of the LDE can be cast as a 3-vector equation in ordinary time as follows. From Eq. (17), one has

$$R_\nu{}^\mu a_\mu(\tau) = \frac{1}{\tau_0 m_0} \int_\tau^\infty d\tau' e^{\frac{\tau-\tau'}{\tau_0}} R_\nu{}^\mu(\tau') f_\mu(\tau'), \tag{31}$$

the left hand side of which is

$$R_\nu{}^\mu a_\mu = \left\{0, \mathbf{u}\frac{du_0}{d\tau} - u_0 \frac{d\mathbf{u}}{d\tau}\right\} = \left\{0, -u_0^2 \frac{d(\mathbf{u}/u_0)}{d\tau}\right\} = \left\{0, -\gamma^3 \frac{d\boldsymbol{\beta}}{dt}\right\}, \tag{32}$$

where $\boldsymbol{\beta} = d\mathbf{x}/dt$ is the ordinary velocity. I.E., the left hand side of Eq. (31) is already in the direction of the ordinary acceleration. Further, noting that the product in the integrand is

$$R_\nu{}^\mu f_\mu = \left\{0, \gamma^2 \left(\boldsymbol{\beta}(\boldsymbol{\beta}.\mathbf{f}) - \mathbf{f}\right)\right\}, \tag{33}$$

then substitution of Eqs. (32) and (33) into Eq. (17) gives

$$\dot{\boldsymbol{\beta}} = \frac{1}{\tau_0 m_0 \gamma^3} \int_\tau^\infty d\tau' e^{\frac{\tau-\tau'}{\tau_0}} \gamma^2 \left(\mathbf{f} - \boldsymbol{\beta}(\boldsymbol{\beta}.\mathbf{f})\right) = \frac{1}{\tau_0 m_0 \gamma^3} \int_t^\infty dt' e^{\frac{\tau-\tau'}{\tau_0}} \mathbf{H}(t'), \tag{34}$$

where the components of $\mathbf{H}(t') = \gamma\left(\mathbf{f} - \boldsymbol{\beta}(\boldsymbol{\beta}.\mathbf{f})\right)$ are now redefined as functions of ordinary time. The transformation is complete once the exponential damping factor is explicitly cast as a function of ordinary time:

$$\dot{\boldsymbol{\beta}} = \frac{1}{\tau_0 m_0 \gamma^3} \int_t^\infty dt' \exp\left(\int_{t'}^t \frac{dt''}{\tau_0 \gamma(t'')}\right) \mathbf{H}(t') = \frac{1}{\tau_0 m_0 \gamma^3} \int_0^\infty dt' \exp\left(-\int_t^{t+t'} \frac{dt''}{\tau_0 \gamma(t'')}\right) \mathbf{H}(t+t'). \tag{35}$$

As for the proper-time form, the variable of integration can be rendered dimensionless, although here it does not result in a simplification. Letting $s = t'/\tau_0$:

$$\frac{d\boldsymbol{\beta}}{dt} = \frac{1}{m_0 \gamma^3} \int_0^\infty ds \exp\left(-\int_t^{t+\tau_0 s} \frac{dt''}{\tau_0 \gamma(t'')}\right) \mathbf{H}(t+\tau_0 s) = \frac{1}{m_0 \gamma^3} \int_0^\infty ds \exp\left(-\int_0^s \frac{ds'}{\gamma(t+\tau_0 s')}\right) \mathbf{H}(t+\tau_0 s). \tag{36}$$

If $\mathbf{f}$ is the Lorentz force then $\mathbf{H} = e\gamma\left(\mathbf{E} - \boldsymbol{\beta}(\boldsymbol{\beta}.\mathbf{E}) + \boldsymbol{\beta} \times \mathbf{B}\right)$.

**Ordinary-time series expansion in $\tau_0$**

An ordinary-time series expansion of the integrand in ascending powers of $\tau_0$ can obtained from Eq. (36) by integrating by parts. The result is

$$\frac{d\boldsymbol{\beta}}{dt} = \frac{1}{m_0 \gamma^3} \sum_{n=0} \left( \gamma \tau_0 \frac{d}{dt} \right)^n \left( \gamma^2 \left( \mathbf{f} - \boldsymbol{\beta}(\boldsymbol{\beta}.\mathbf{f}) \right) \right), \tag{37}$$

where the functions are of ordinary time, evaluated at time $t$.

**Summary**

A physically acceptable relativistic equation of motion for a classical charged particle in 3 spatial dimensions has been derived that has the properties desired of the original Lorentz-Dirac equation, but without the unphysical behavior. The exclusion of runaway solutions has been achieved by finding an integrating factor for the original Lorentz-Dirac equation so that the acceleration can be written as an integral operator on the force.